\documentclass[journal]{IEEEtran}
\usepackage{mathtools}

\usepackage{fancyhdr}
\usepackage{amsmath,amssymb}
\usepackage{amsfonts}
\usepackage{graphicx}
\usepackage{dsfont}
\usepackage{comment}
\usepackage{algorithmic}
\usepackage{comment}
\usepackage{caption}
\usepackage{alphabeta}
\usepackage{balance}
\usepackage{algorithm}
\usepackage{float}
\usepackage[hidelinks]{hyperref}    

\usepackage{color}
\usepackage{pgfplots}
\pgfplotsset{compat=1.18}
\usepackage{subcaption}

\usepackage{enumerate}
\usepackage{graphics}
\usepackage{cite}
\usepackage{mathrsfs}
\usepackage{stfloats}

\usepackage{tikz}
\usepackage{mathdots}
\usepackage{yhmath}
\usepackage{cancel}
\usepackage{color}
\usepackage{siunitx}
\usepackage{array}
\usepackage{multirow}
\usepackage{textcomp}
\usepackage{gensymb}
\usepackage{tabularx}
\usepackage{booktabs}
\usetikzlibrary{fadings}
\usetikzlibrary{patterns}
\usetikzlibrary{shadows.blur}
\usepackage{graphicx}

\begin{document}

\title{Jacobi Elliptic Chirps for Sub-Nyquist \\ Multi-Target Ranging}
\author{Dimitrios Bozanis, Thrassos K. Oikonomou, Sotiris A. Tegos,~\IEEEmembership{Senior Member, IEEE},\\ Panagiotis D. Diamantoulakis,~\IEEEmembership{Senior Member, IEEE}, and George K. Karagiannidis,~\IEEEmembership{Fellow, IEEE}
\thanks{D. Bozanis, T. K. Oikonomou, S. A. Tegos, P. D. Diamantoulakis, and G. K. Karagiannidis are with the Department of Electrical and Computer Engineering, Aristotle University of Thessaloniki, Thessaloniki, Greece (e-mails: \{dimimpoz, toikonom, tegosoti, padiaman, geokarag\}@ece.auth.gr)}
}
\maketitle

\begin{abstract} 
Sub-Nyquist sampling is an attractive way to reduce the hardware cost of wideband pulse-compression radar, but it introduces coherent alias-induced replicas in the matched-filter range profile, producing spurious peaks known as ghost targets. Existing frequency-modulated waveforms face a practical trade-off in this regime: linear frequency-modulated (LFM) pulses provide compact range responses but are highly susceptible to ghost-target detections, whereas hyperbolic frequency-modulated (HFM) pulses suppress ghosts at the cost of degraded target separability. To overcome this trade-off, we propose a sine-over-cosine Jacobi elliptic frequency-modulated waveform, referred to as SC-EFM, in which the elliptic modulus tunes the instantaneous-frequency (IF) curvature while preserving the pulse duration and bandwidth of conventional benchmarks. We characterize the sub-Nyquist folding structure of SC-EFM and derive closed-form expressions for the multi-target and ghost-target detection probabilities. Numerical results show that SC-EFM significantly suppresses ghost detections relative to LFM while matching its target separability, and substantially outperforms HFM in resolving close targets, providing a unified waveform solution for ghost-resilient sub-Nyquist multi-target ranging.
\end{abstract}


\section{Introduction}

Wireless sensing is becoming a core capability of future connected systems, where communication nodes are expected to provide reliable environmental awareness in addition to data exchange~\cite{Liu,Zhang}. In vehicular, infrastructure-assisted, and integrated sensing and communication scenarios, pulse-compression radar is particularly attractive because it enables high range resolution by transmitting long intra-pulse-modulated wideband signals and compressing the received echoes at the receiver. Linear frequency-modulated (LFM) pulses remain a standard choice due to their simple generation, compact main lobe, and well-understood range response~\cite{chirp}. However, wideband pulse-compression radar requires high-rate analog-to-digital converters, large memory bandwidth, and high processing throughput. These requirements become increasingly restrictive in low-cost sensing nodes when the transmitted bandwidth is large and the pulse duration is short.

Several approaches have been studied to reduce the sampling-rate requirements of wideband radar receivers. In pulse-compression radar, increasing the bandwidth improves range resolution, but it also raises the nominal Nyquist rate, and therefore the ADC, memory, and processing burden. Sub-Nyquist sampling is thus attractive for reducing receiver complexity while retaining wideband sensing capabilities. Existing solutions include compressive sensing, Xampling, Nyquist-folding architectures, and waveform-dependent sub-Nyquist recovery methods~\cite{Candes_CS_2006,Herman_CS_Radar_2009,Mishali_Xampling_2011,Fudge_NyquistFolding_2008,Cohen_SubNyquistRadar_2018}. However, in matched-filter pulse-compression radar, direct sub-Nyquist sampling of LFM pulses is problematic because frequency folding and aliasing may generate coherent range-profile replicas, which appear as ghost targets. Recent work has shown that hyperbolic frequency-modulated (HFM) pulses can mitigate this effect by producing nonuniform folded and aliased components~\cite{HFM}. Nevertheless, HFM waveforms may exhibit a compressed response that is broader than that of LFM, which can degrade the separability of closely spaced targets.

This letter is motivated by the resulting waveform design trade-off. LFM provides a compact range response but suffers from structured alias ghosts under sub-Nyquist sampling, while HFM suppresses such ghosts at the cost of reduced target separability. A waveform with tunable nonlinear curvature can therefore bridge these two regimes. To the best of our knowledge, Jacobi-elliptic-function-based frequency modulation has not been investigated as a tunable waveform family for ghost-resilient sub-Nyquist pulse-compression ranging.

The contribution of this letter is threefold. First, we introduce a sine-over-cosine Jacobi elliptic frequency-modulated pulse, referred to as SC-EFM, whose elliptic modulus controls the instantaneous-frequency (IF) curvature while preserving the same pulse duration and bandwidth as LFM and HFM. Second, we derive the sub-Nyquist folding structure of SC-EFM and show that the folding-boundary crossing times are analytically controlled by the elliptic modulus, providing a mechanism to reshape alias-induced correlations. Third, we derive closed-form expressions for the multi-target and ghost-target detection probabilities under sub-Nyquist sampling. Numerical results demonstrate that SC-EFM suppresses ghost targets compared to LFM while maintaining target separability significantly better than HFM.


\section{System Model}
\label{sec:system_model}

We consider a monostatic pulse-compression radar that illuminates $L$ static targets and transmits a train of $P$ identical frequency-modulated pulses with duration $T_p$, bandwidth $B=f_H-f_L$, and pulse repetition interval (PRI) $T_r$. Hence, the proposed unit-energy SC-EFM pulse is written as
\begin{equation}
s_1(t)=\frac{1}{\sqrt{T_p}}
\operatorname{rect}\left(\frac{t-T_p/2}{T_p}\right)
\exp\left(j2\pi \varphi_E(t)\right) ,
\label{eq:pulse}
\end{equation}
where
\begin{equation}
\operatorname{rect}(x)
=
\begin{cases}
1, & |x|<\frac{1}{2},\\
0, & \text{otherwise}
\end{cases}
\label{eq:rect}
\end{equation}
and
\begin{equation}
\varphi_E(t)=\int_0^t f_E(\tau)d\tau .
\label{eq:phase}
\end{equation}
The IF is defined over $0\leq t\leq T_p$ as
\begin{equation}
f_E(t;m)=f_L+B g_E(t;m),
\end{equation}
where the semicolon separates the time variable $t$ from the waveform-shaping parameter $m$, and
\begin{equation}
g_E(t;m)
=
\frac{\operatorname{sn}(u(t);m)}
{1+\operatorname{cn}(u(t);m)},
\qquad
u(t)=\frac{K(m)}{T_p}t .
\label{eq:sc_efm_law}
\end{equation}
Here, $\operatorname{sn}(\cdot;m)$ and $\operatorname{cn}(\cdot;m)$ are Jacobi elliptic functions, $K(m)$ is the complete elliptic integral of the first kind~\cite{ByrdFriedman}, and $m\in(0,1)$ is the elliptic modulus. Since $\operatorname{sn}(0;m)=0$, $\operatorname{cn}(0;m)=1$, $\operatorname{sn}(K(m);m)=1$, and $\operatorname{cn}(K(m);m)=0$, the normalized sweep satisfies $
g_E(0;m)=0,
\:
g_E(T_p;m)=1.
$
Thus, SC-EFM sweeps over the frequency interval $[f_L,f_H]$ with bandwidth $B$.

The transmitted pulse train is
\begin{equation}
s(t)=\sum_{p=0}^{P-1}s_1(t-pT_r).
\label{eq:pulse_train}
\end{equation}
The $l$-th target is characterized by range $R_l$, round-trip delay $\tau_l=2R_l/c,$, and complex reflection coefficient $\alpha_l$, where $c$ is the propagation speed. Therefore, the received baseband signal is
\begin{equation}
x(t)
=
\sum_{p=0}^{P-1}
\sum_{l=0}^{L-1}
\alpha_l s_1(t-\tau_l-pT_r)
+
w(t),
\label{eq:rx_signal_global}
\end{equation}
where $w(t)$ is circularly symmetric complex Gaussian noise with zero mean and variance $\sigma_w^2$.

The received signal is divided into PRIs. The samples associated with the $p$-th pulse are then expressed as
\begin{equation}
x_p(t)
=
x(t+pT_r)
=
\sum_{l=0}^{L-1}
\alpha_l s_1(t-\tau_l)
+
w_p(t),
\quad
0\leq t<T_r,
\label{eq:rx_signal}
\end{equation}
where $w_p(t)=w(t+pT_r)$ denotes the noise component in the $p$-th PRI.
Each received pulse is then uniformly sampled at the sub-Nyquist rate $f_s'=\eta f_{\rm nyq},
\:
0<\eta\leq1,$
where $f_{\rm nyq}=2B$ is the nominal Nyquist rate and $\eta$ is the sampling factor. The sampling interval is $T_s'=\frac{1}{f_s'}.$ Let $n$ denote the fast-time sample index. Hence, the sampled received pulse is correlated with the sampled reference pulse, yielding the matched-filter range profile
\begin{equation}
y_p[q]
=
\sum_n
x_p(nT_s')s_1^*(nT_s'-qT_s'),
\label{eq:matched_filter}
\end{equation}
where $q$ denotes the delay-bin index. A delay peak at $q_l$ corresponds to the range estimate
\begin{equation}
\hat R_l=\frac{c}{2}\frac{q_l}{f_s'}.
\label{eq:range_estimate}
\end{equation}

\section{Sub-Nyquist Sampling of SC-EFM}
\label{sec:subnyquist_sampling}

Let $F=f_s'/2$ denote the folding frequency. After sampling at $f_s'$, the time-varying IF is periodically folded into the observable interval $[0,F]$. The folded IF of SC-EFM can therefore be written as
\begin{equation}
\widetilde f_E(t)
=
\left|
\operatorname{mod}\left(f_E(t;m)+F,f_s'\right)-F
\right|.
\label{eq:folded_frequency}
\end{equation}
This expression describes the frequency folding induced by sampling at $f_s'$.
The folding structure is determined by the instants at which the original IF crosses integer multiples of the folding boundary. These crossing times satisfy
\begin{equation}
f_E(t_k;m)=kF,
\label{eq:folding_boundary}
\end{equation}
or equivalently
\begin{equation}
g_E(t_k;m)=\xi_k,
\qquad
\xi_k=\frac{kF-f_L}{B},
\label{eq:xi_definition}
\end{equation}
for all integers $k$ such that $0\leq \xi_k\leq1$. Using the definition of $g_E(t;m)$ in \eqref{eq:sc_efm_law}, the crossing equation becomes
\begin{equation}
\frac{\operatorname{sn}(u_k;m)}
{1+\operatorname{cn}(u_k;m)}
=
\xi_k,
\qquad
u_k=u(t_k)=\frac{K(m)}{T_p}t_k .
\label{eq:efm_crossing_equation}
\end{equation}

Let $ 0=b_0<b_1<\cdots<b_{N_E}=T_p $ denote the ordered set formed by the pulse endpoints and all valid folding-boundary crossing times $t_k$, where $N_E$ is the resulting number of folded segments. The $i$-th folded segment occupies the interval $[b_{i-1},b_i)$ and has duration \begin{equation} \Delta T_i^E=b_i-b_{i-1}, \qquad i=1,\ldots,N_E. \label{eq:efm_segment_duration} \end{equation}
The crossing equation in \eqref{eq:efm_crossing_equation} shows that the ordered boundaries $b_i$, and therefore the segment durations $\Delta T_i^E$, are directly controlled by the elliptic modulus $m$.
Since differentiating $g_E(t;m)$ yields a positive derivative over $0\leq u(t)\leq K(m)$, with $\operatorname{dn}(u(t);m)>0$ and $\operatorname{cn}(u(t);m)\in[0,1]$, the SC-EFM sweep is monotonic over the pulse duration, and each valid folding-boundary level has a unique crossing time.

The segment durations in \eqref{eq:efm_segment_duration} reveal the central design freedom of SC-EFM. For LFM, the affine IF leads to nearly equal crossing intervals, except for the first and last partial segments. These equal intervals can make the folded and aliased portions highly coherent, producing ghost targets at range bins displaced from the true target locations. For HFM, the nonlinear hyperbolic law produces unequal crossing intervals, which suppresses coherent alias replicas~\cite{HFM}. Similarly, SC-EFM produces unequal crossing intervals, but these intervals are controlled by the elliptic modulus $m$. Therefore, SC-EFM provides an additional degree of freedom for controlling the alias structure and the compressed range response. This effect is reflected in the off-zero autocorrelation values of the sampled waveform, which determine the sidelobe and ghost-target behavior of the matched-filter range profile.

\input{figure1}

Fig.~\ref{fig:if_tikz_subfigures} illustrates the time-frequency signatures of LFM, HFM, and the proposed SC-EFM under Nyquist and sub-Nyquist sampling. LFM produces folded portions with nearly uniform time spacing, whereas HFM produces nonuniform folded portions due to its nonlinear frequency law. SC-EFM occupies a tunable intermediate regime: its unfolded IF curve remains a smooth sweep over the same bandwidth, while its folded counterpart exhibits nonuniform crossing intervals controlled by $m$. This supports the use of SC-EFM for combining HFM-like ghost suppression with a compact compressed range response.

\section{Received Signal Processing}
\label{sec:rx_processing}

Building on the matched-filter range profile in \eqref{eq:matched_filter}, this section characterizes the range-processing output of the proposed SC-EFM waveform and derives the corresponding true-target and ghost-target detection probabilities. 

\subsection{Range-Profile Model}

Using the received signal model in \eqref{eq:rx_signal}, the range profile of the $p$-th received pulse can be written as
\begin{equation}
y_p[q]
=
\sum_{l=0}^{L-1}
\alpha_l
R_E\left(qT_s'-\tau_l\right)
+
\nu_p[q],
\label{eq:range_profile_model_tau}
\end{equation}
where
\begin{equation}
R_E(\Delta)
=
\sum_n
s_1(nT_s')
s_1^*(nT_s'-\Delta)
\label{eq:discrete_autocorrelation}
\end{equation}
is the discrete autocorrelation of the sub-Nyquist sampled SC-EFM pulse, and
\begin{equation}
\nu_p[q]
=
\sum_n
w_p(nT_s')s_1^*(nT_s'-qT_s')
\label{eq:post_correlation_noise}
\end{equation}
is the noise component at the output of the matched filter.
Without loss of generality, we assume that the $l$-th delay is aligned with the sampling grid, such that $q_l=\tau_l f_s'$ and \eqref{eq:range_profile_model_tau} becomes
\begin{equation}
y_p[q]
=
\sum_{l=0}^{L-1}
\alpha_l
R_E\left((q-q_l)T_s'\right)
+
\nu_p[q].
\label{eq:range_profile_model}
\end{equation}

From the noise model in Section~\ref{sec:system_model}, the matched-filter noise term satisfies $
\nu_p[q]\sim\mathcal{CN}(0,\sigma_\nu^2),
\:
\sigma_\nu^2=\sigma_w^2 E_d,
$
where
\begin{equation}
E_d
=
\sum_n |s_1(nT_s')|^2
\label{eq:discrete_energy}
\end{equation}
is the discrete energy of the sampled reference pulse used in the correlation, and $\sigma_w^2$ is the per-sample noise variance. Accordingly, the matched-filter output SNR is defined as
$
\mathrm{SNR}_{\mathrm{MF}}
\triangleq
|\alpha E_d|^2/\sigma_\nu^2
=
|\alpha|^2E_d/\sigma_w^2
$. Normalizing the discrete autocorrelation in \eqref{eq:discrete_autocorrelation} by the reference energy $E_d$ and evaluating it at delay lags $\Delta=\ell T_s'$, we define
\begin{equation}
\chi_E[\ell]
=
\frac{1}{E_d}
\sum_n
s_1(nT_s')
s_1^*(nT_s'-\ell T_s'),
\label{eq:normalized_corr}
\end{equation}
where $\ell$ is an integer delay-lag index. By construction, $\chi_E[0]=1$. Using \eqref{eq:normalized_corr}, \eqref{eq:range_profile_model} can be equivalently expressed as
\begin{equation}
y_p[q]
=
\sum_{l=0}^{L-1}
\alpha_l E_d
\chi_E[q-q_l]
+
\nu_p[q].
\label{eq:range_profile_normalized}
\end{equation}
The off-target values of $\chi_E[\ell]$, with $\ell\neq0$, determine the sidelobe and ghost-target structure of the range profile. Under sub-Nyquist sampling, these off-zero correlation values are directly affected by the frequency-folding structure of the waveform. Thus, the SC-EFM folding intervals controlled by $m$ can reshape the alias-induced correlation terms that appear after matched filtering. From \eqref{eq:range_profile_normalized}, the deterministic component of the range profile at an arbitrary delay bin $q$ is
\begin{equation}
\mu[q]
=
\mathbb{E}\{y[q]\}
=
\sum_{l=0}^{L-1}
\alpha_l E_d
\chi_E[q-q_l].
\label{eq:deterministic_profile}
\end{equation}
Therefore, reducing the off-zero values of $|\chi_E[\ell]|$ decreases the probability that an off-target delay bin produces a large correlation peak and is detected as a ghost target.
In the following detection analysis, we use the single-pulse matched-filter output and write $y[q]=y_p[q]$ for notational simplicity.
\subsection{Multi-Target Detection Probability}

We next consider target detection in a multi-target scenario. For the $k$-th target at delay bin $q_k$, the contribution of all targets except the $k$-th one is obtained from \eqref{eq:deterministic_profile} as
\begin{equation}
\mu_{-k}[q_k]
=
\sum_{\substack{l=0\\ l\neq k}}^{L-1}
\alpha_l E_d
\chi_E[q_k-q_l].
\label{eq:multi_target_leakage}
\end{equation}
Thus, the matched-filter output at $q_k$ can be written under the two hypotheses as
\begin{align}
\mathcal{H}_{0,k} &: \quad
y[q_k]
=
\mu_{-k}[q_k]
+
\nu[q_k],
\\
\mathcal{H}_{1,k} &: \quad
y[q_k]
=
\mu_{-k}[q_k]
+
\alpha_k E_d
+
\nu[q_k].
\end{align}
Here, $\mathcal{H}_{0,k}$ does not correspond to a noise-only bin. Instead, it represents the case where the remaining targets are present, but the $k$-th target is absent. 

We employ the magnitude detector $z[q]=|y[q]|$. Thus, for the $k$-th target, the decision rule is
\begin{equation}
z[q_k]>\gamma_k \ \Rightarrow \ \mathcal{H}_{1,k},
\qquad
z[q_k]\leq\gamma_k \ \Rightarrow \ \mathcal{H}_{0,k},
\label{eq:multi_target_detector}
\end{equation}
where $\gamma_k$ is the detection threshold. Since $y[q_k]$ under $\mathcal{H}_{0,k}$ is complex Gaussian with nonzero mean $\mu_{-k}[q_k]$, $z[q_k]$ follows a Rician distribution. Therefore, the false-alarm probability is
\begin{equation}
P_{\mathrm{FA},k}
=
Q_1
\left(
\frac{\sqrt{2}|\mu_{-k}[q_k]|}{\sigma_\nu},
\frac{\sqrt{2}\gamma_k}{\sigma_\nu}
\right),
\label{eq:multi_target_pfa}
\end{equation}
where $Q_1(\cdot,\cdot)$ is the first-order Marcum $Q$-function. If the leakage term is negligible, i.e., $\mu_{-k}[q_k]=0$, the null hypothesis becomes $y[q_k]=\nu[q_k]$. In this case, $z[q_k]=|y[q_k]|$ is Rayleigh distributed, and \eqref{eq:multi_target_pfa} reduces to the noise-only threshold
\begin{equation}
    \gamma_k
    =
    \sigma_\nu\sqrt{-\ln P_{\mathrm{FA},k}}.
    \label{eq:threshold}
\end{equation}

Under $\mathcal{H}_{1,k}$, the deterministic component becomes $\mu_{-k}[q_k]+\alpha_k E_d$. Thus, the detection probability of the $k$-th target is
\begin{equation}
P_{\mathrm{D},k}
=
Q_1
\left(
\frac{
\sqrt{2}
\left|
\mu_{-k}[q_k]+\alpha_k E_d
\right|
}{\sigma_\nu},
\frac{\sqrt{2}\gamma_k}{\sigma_\nu}
\right),
\label{eq:multi_target_pd}
\end{equation}
which shows that multi-target detection is waveform-dependent through the off-zero autocorrelation values $\chi_E[q_k-q_l]$ appearing in \eqref{eq:multi_target_leakage}. These values quantify how strongly the remaining targets leak into the delay bin of interest.

\subsection{Ghost-Target Detection Probability}

We now characterize the probability that an off-target delay bin is declared as a target due to waveform sidelobes or alias-induced correlation. Consider a candidate ghost bin $q_g$, where $q_g\notin\mathcal{Q}$, with $
\mathcal{Q}=\{q_0,q_1,\ldots,q_{L-1}\}$
denoting the set of true target bins. From \eqref{eq:range_profile_normalized}, the matched-filter output at $q_g$ is
\begin{equation}
y[q_g]
=
\mu_g
+
\nu[q_g],
\label{eq:ghost_bin_output}
\end{equation}
where
\begin{equation}
\mu_g
=
\sum_{l=0}^{L-1}
\alpha_l E_d
\chi_E[q_g-q_l].
\label{eq:ghost_mu}
\end{equation}
Although $q_g$ does not correspond to a physical target, the detector observes a nonzero deterministic component whenever the waveform correlation at that delay is non-negligible.

The ghost-bin hypotheses can be written as
\begin{align}
\mathcal{G}_0 &: \quad y[q_g]=\nu[q_g],\\
\mathcal{G}_1 &: \quad y[q_g]=\mu_g+\nu[q_g].
\end{align}
Here, $\mathcal{G}_1$ does not indicate the presence of a real target. Instead, it represents a waveform-induced false peak due to sidelobe or alias leakage. The ghost-target detection probability is therefore
\begin{equation}
P_{\mathrm{G}}(q_g)
=
\Pr\{|y[q_g]|>\gamma \mid \mu_g\}.
\label{eq:ghost_probability_definition}
\end{equation}
Since $y[q_g]$ is complex Gaussian with mean $\mu_g$ and variance $\sigma_\nu^2$, $|y[q_g]|$ is Rician distributed, thus
\begin{equation}
P_{\mathrm{G}}(q_g)
=
Q_1
\left(
\frac{\sqrt{2}|\mu_g|}{\sigma_\nu},
\frac{\sqrt{2}\gamma}{\sigma_\nu}
\right).
\label{eq:pg_bin}
\end{equation}
Using the noise-only threshold in \eqref{eq:threshold}, we obtain
\begin{equation}
P_{\mathrm{G}}(q_g)
=
Q_1
\left(
\frac{\sqrt{2}|\mu_g|}{\sigma_\nu},
\sqrt{-2\ln P_{\mathrm{FA}}}
\right),
\label{eq:pg_final}
\end{equation}
which shows that the threshold is determined by the prescribed noise-only false-alarm probability and the post-correlation noise variance, whereas the ghost-target probability is waveform-dependent through $\mu_g$. Thus, reducing the off-zero correlation values $\chi_E[\ell]$ reduces the deterministic leakage term $\mu_g$ and, consequently, the probability of ghost detection.

\section{Numerical Results}

In this section, we numerically evaluate the range-processing performance of the proposed SC-EFM waveform and compare it with the conventional LFM and HFM benchmarks. All waveforms are evaluated under the same pulse duration, bandwidth, and sub-Nyquist sampling factor. The main simulation parameters are summarized in Table~\ref{tab:simulation_parameters} \cite{HFM}. Moreover, solid and dashed curves denote the theoretical closed-form results, while square and circle markers denote Monte Carlo (MC) simulations.

\begin{table}[t]
\centering
\caption{Simulation parameters.}
\label{tab:simulation_parameters}
\begin{tabular}{l c}
\hline
Parameter & Value \\
\hline
Lower sweep frequency, $f_L$ & $50~\mathrm{MHz}$ \\
Bandwidth, $B$ & $300~\mathrm{MHz}$ \\
Pulse duration, $T_p$ & $2~\mu\mathrm{s}$ \\
Pulse repetition interval, $T_r$ & $10~\mu\mathrm{s}$ \\
Sampling factor, $\eta$ & $0.4$ \\
SC-EFM modulus, $m$ & $0.75$ \\
Number of targets, $L$ & $2$ \\
Reference target range & $20~\mathrm{m}$ \\
Target separation, $\Delta R$ & $1.4~\mathrm{m}$ \\
False-alarm probability, $P_{\mathrm{FA}}$ & $10^{-3}$ \\
\hline
\end{tabular}
\end{table}

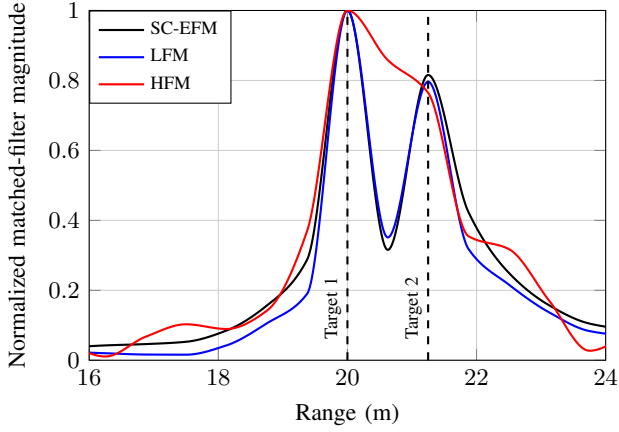
\begin{figure}[t]
\centering
\begin{tikzpicture}

\begin{axis}[
    width=0.95\columnwidth,
    height=6.2cm,
    xmin=16, xmax=24.0,
    ymin=0, ymax=1.00,
    grid=both,
    grid style={line width=.1pt, draw=gray!25},
    major grid style={line width=.2pt, draw=gray!40},
    xlabel={\small Range (m)},
    ylabel={\small Normalized matched-filter magnitude},
    tick label style={font=\small},
    label style={font=\small},
    title style={font=\small\bfseries},
    legend style={
        font=\scriptsize,
        draw=black,
        fill=white,
        at={(0,1)},
        anchor=north west
    },
    legend cell align={left},
]


\addplot[
    black,
    thick
]
table[
    x=R_m,
    y=SC_EFM_norm,
    col sep=tab
]{fig2_close_target_separability_curves.dat};
\addlegendentry{SC-EFM}

\addplot[
    blue,
    thick
]
table[
    x=R_m,
    y=LFM_norm,
    col sep=tab
]{fig2_close_target_separability_curves.dat};
\addlegendentry{LFM}

\addplot[
    red,
    thick
]
table[
    x=R_m,
    y=HFM_norm,
    col sep=tab
]{fig2_close_target_separability_curves.dat};
\addlegendentry{HFM}


\addplot[
    black,
    dashed,
    thick
]
coordinates {(20,0) (20,1.05)};

\addplot[
    black,
    dashed,
    thick
]
coordinates {(21.25,0) (21.25,1.05)};

\node[rotate=90, anchor=south, font=\scriptsize] at (axis cs:20.05,0.15) {Target 1};
\node[rotate=90, anchor=south, font=\scriptsize] at (axis cs:21.3,0.15) {Target 2};

\end{axis}

\end{tikzpicture}
\caption{\small Normalized matched-filter range profile.}
\label{fig:range_profiles}
\end{figure}

Fig.~\ref{fig:range_profiles} shows the normalized matched-filter range profiles for two closely spaced targets. The HFM waveform produces a broad response around the two target positions and does not clearly resolve the two echoes. In particular, the two target responses strongly overlap for HFM, resulting in a single dominant range peak rather than two clearly separated peaks. In contrast, both LFM and SC-EFM provide two distinguishable peaks close to the true target ranges. This result shows that the proposed SC-EFM waveform preserves close-target separability to a large extent, while avoiding the severe peak merging observed for HFM.

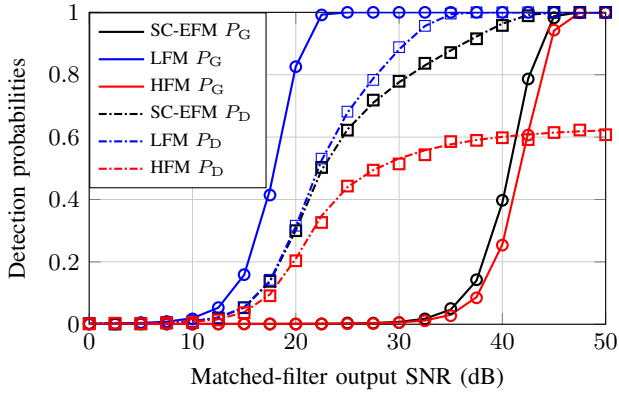
\begin{figure}[t]
\centering
\begin{tikzpicture}

\begin{axis}[
    width=0.95\columnwidth,
    height=5.7cm,
    xmin=0, xmax=50,
    ymin=0, ymax=1.0,
    xtick={0,10,20,30,40,50},
    ytick={0,0.2,0.4,0.6,0.8,1},
    grid=both,
    grid style={line width=.1pt, draw=gray!25},
    major grid style={line width=.2pt, draw=gray!40},
    xlabel={Matched-filter output SNR (dB)},
    ylabel={Detection probabilities},
    tick label style={font=\small},
    label style={font=\small},
    title style={font=\small\bfseries},
    legend style={
        font=\scriptsize,
        draw=black,
        fill=white,
        at={(0,1)},
        anchor=north west
    },
    legend cell align={left},
]

\addplot[
    black,
    thick
]
table[
    x=SNR_dB,
    y=HFM_theory,
    col sep=tab
]{fig3_ghost_probability_eta_0p30.dat};
\addlegendentry{SC-EFM $P_{\mathrm{G}}$}

\addplot[
    only marks,
    mark=o,
    mark size=2pt,
    black,
    thick,
    forget plot
]
table[
    x=SNR_dB,
    y=HFM_MC,
    col sep=tab
]{fig3_ghost_probability_eta_0p30.dat};


\addplot[
    blue,
    thick
]
table[
    x=SNR_dB,
    y=LFM_theory,
    col sep=tab
]{fig3_ghost_probability_eta_0p30.dat};
\addlegendentry{LFM $P_{\mathrm{G}}$}

\addplot[
    only marks,
    mark=o,
    mark size=2pt,
    blue,
    thick,
    forget plot
]
table[
    x=SNR_dB,
    y=LFM_MC,
    col sep=tab
]{fig3_ghost_probability_eta_0p30.dat};


\addplot[
    red,
    thick
]
table[
    x=SNR_dB,
    y=SC_EFM_theory,
    col sep=tab
]{fig3_ghost_probability_eta_0p30.dat};
\addlegendentry{HFM $P_{\mathrm{G}}$}

\addplot[
    only marks,
    mark=o,
    mark size=2pt,
    red,
    thick,
    forget plot
]
table[
    x=SNR_dB,
    y=SC_EFM_MC,
    col sep=tab
]{fig3_ghost_probability_eta_0p30.dat};

\addplot[
    black,
    thick,
    densely dashdotted
]
table[
    x=SNR_dB,
    y=SC_EFM_theory,
    col sep=tab
]{fig5b_weak_detection_snr_eta_0p40_sep_1p40.dat};
\addlegendentry{SC-EFM $P_{\mathrm{D}}$}

\addplot[
    only marks,
    mark=square,
    mark size=2pt,
    black,
    thick,
    forget plot
]
table[
    x=SNR_dB,
    y=SC_EFM_MC,
    col sep=tab
]{fig5b_weak_detection_snr_eta_0p40_sep_1p40.dat};


\addplot[
    blue,
    thick,
    densely dashdotted
]
table[
    x=SNR_dB,
    y=LFM_theory,
    col sep=tab
]{fig5b_weak_detection_snr_eta_0p40_sep_1p40.dat};
\addlegendentry{LFM $P_{\mathrm{D}}$}

\addplot[
    only marks,
    mark=square,
    mark size=2pt,
    blue,
    forget plot
]
table[
    x=SNR_dB,
    y=LFM_MC,
    col sep=tab
]{fig5b_weak_detection_snr_eta_0p40_sep_1p40.dat};


\addplot[
    red,
    thick,
    densely dashdotted
]
table[
    x=SNR_dB,
    y=HFM_theory,
    col sep=tab
]{fig5b_weak_detection_snr_eta_0p40_sep_1p40.dat};
\addlegendentry{HFM $P_{\mathrm{D}}$}

\addplot[
    only marks,
    mark=square,
    mark size=2pt,
    red,
    thick,
    forget plot
]
table[
    x=SNR_dB,
    y=HFM_MC,
    col sep=tab
]{fig5b_weak_detection_snr_eta_0p40_sep_1p40.dat};

\end{axis}

\end{tikzpicture}
\caption{\small Target and ghost detection probabilities vs. $\mathrm{SNR}_\mathrm{MF}$.}
\label{fig:combined_detection_ghost}
\end{figure}

Fig.~\ref{fig:combined_detection_ghost} reports the ghost detection probability and the target detection probability versus the matched-filter output SNR, $\mathrm{SNR}_\mathrm{MF}$. The LFM waveform achieves high target detection probability at relatively low SNR, but its ghost probability also increases rapidly. This indicates that LFM is highly sensitive to the alias-induced correlation created by sub-Nyquist sampling. On the other hand, HFM provides lower ghost probability over a wide SNR range, but this comes at the cost of degraded target detection, which is consistent with its poor separability in Fig.~\ref{fig:range_profiles}. In contrast, the proposed SC-EFM waveform achieves a more favorable balance: it substantially suppresses ghost detections compared with LFM, while maintaining much better target detection performance than HFM. This highlights the main advantage of the proposed design, namely that the elliptic shaping of the folded IF curve can jointly improve ghost suppression and target separability. Finally, the close agreement between the curves and the MC markers verifies the analytical expressions.

\begin{figure}[t]
\centering
\begin{tikzpicture}

\begin{axis}[
    width=0.95\columnwidth,
    height=5.8cm,
    xmin=0.5, xmax=10,
    ymin=0, ymax=1,
    xtick={1,2,...,10},
    ytick={0,0.2,0.4,0.6,0.8,1},
    grid=both,
    grid style={line width=.1pt, draw=gray!25},
    major grid style={line width=.2pt, draw=gray!40},
    xlabel={Target separation $\Delta R$ (m)},
    ylabel={Target detection probability},
    tick label style={font=\small},
    label style={font=\small},
    legend style={
        font=\scriptsize,
        draw=black,
        fill=white,
        at={(0,0)},
        anchor=south west
    },
    legend cell align={left},
]


\addplot[
    black,
    thick
]
table[
    x index=0,
    y index=5,
    col sep=tab
]{fig5a_weak_detection_sep_eta_0p40.dat};
\addlegendentry{SC-EFM $P_{\mathrm{D}}$}

\addplot[
    only marks,
    mark=square,
    mark size=1.5pt,
    black,
    thick,
    forget plot
]
table[
    x index=0,
    y index=6,
    col sep=tab
]{fig5a_weak_detection_sep_eta_0p40.dat};


\addplot[
    blue,
    thick
]
table[
    x index=0,
    y index=1,
    col sep=tab
]{fig5a_weak_detection_sep_eta_0p40.dat};
\addlegendentry{LFM $P_{\mathrm{D}}$}

\addplot[
    only marks,
    mark=square,
    mark size=1.5pt,
    blue,
    thick,
    forget plot
]
table[
    x index=0,
    y index=2,
    col sep=tab
]{fig5a_weak_detection_sep_eta_0p40.dat};


\addplot[
    red,
    thick
]
table[
    x index=0,
    y index=3,
    col sep=tab
]{fig5a_weak_detection_sep_eta_0p40.dat};
\addlegendentry{HFM $P_{\mathrm{D}}$}

\addplot[
    only marks,
    mark=square,
    mark size=1.5pt,
    red,
    thick,
    forget plot
]
table[
    x index=0,
    y index=4,
    col sep=tab
]{fig5a_weak_detection_sep_eta_0p40.dat};

\end{axis}

\end{tikzpicture}
\caption{\small Target detection probability vs. $\Delta R$ for $\mathrm{SNR}_\mathrm{MF}=25~\mathrm{dB}$.}
\label{fig:weak_target_detection_separation}
\end{figure}
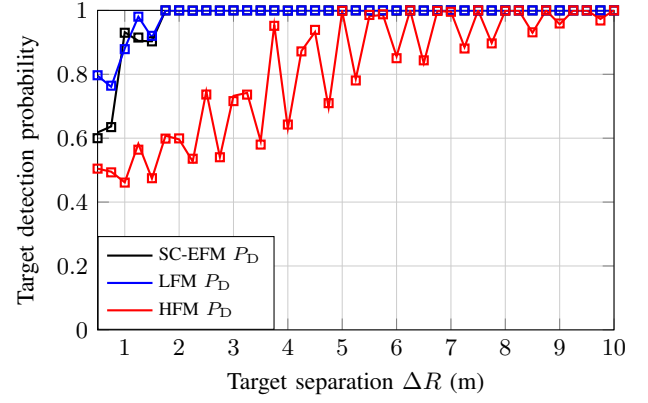

Fig.~\ref{fig:weak_target_detection_separation} shows the target-detection probability as a function of the target separation, with $\mathrm{SNR}_\mathrm{MF}=25~\mathrm{dB}$. For small $\Delta R$, the two range responses overlap, and leakage between the corresponding delay bins reduces the detection probability. As $\Delta R$ increases, the LFM and SC-EFM curves rapidly approach 1, indicating reliable detection once the two responses become sufficiently separated. In contrast, HFM does not reach the same detection level even at larger separations, which is consistent with its broader range response and the persistent inter-target leakage observed in Fig.~\ref{fig:range_profiles}. The oscillatory behavior is due to the separation-dependent autocorrelation terms in \eqref{eq:multi_target_leakage}, which make the inter-target leakage alternate between constructive and destructive contributions. Compared to HFM, SC-EFM maintains a higher detection probability across the entire region while closely matching the LFM benchmark.

\section{Conclusion}

This letter proposed SC-EFM, a Jacobi-elliptic frequency-modulated waveform for sub-Nyquist pulse-compression ranging. By controlling the instantaneous-frequency curvature through the elliptic modulus, SC-EFM reshapes the folding structure induced by sub-Nyquist sampling. Closed-form expressions for target and ghost detection probabilities were derived, showing how the waveform-dependent autocorrelation terms affect multi-target ranging performance. Numerical results demonstrated that SC-EFM suppresses structured ghost detections compared to LFM, while preserving target separability significantly better than HFM. Future work will focus on systematic optimization of the elliptic modulus $m$ under different sub-Nyquist sampling conditions.


\bibliographystyle{IEEEtran}
\bibliography{bibliography}
\end{document}